\documentclass[sigconf, nonacm,balance=false]{acmart}

\newcommand\vldbdoi{XX.XX/XXX.XX}

\newcommand\vldbavailabilityurl{URL_TO_YOUR_ARTIFACTS}
\newcommand\vldbpagestyle{plain}

\newcommand*{\hide}[1]{}

\usepackage{xspace}
\usepackage{subcaption}
\usepackage[frozencache]{minted}
\usepackage{multirow}
\usepackage{tikz}
\usepackage{fontawesome}
\usetikzlibrary{angles,calc,intersections,quotes,arrows.meta,positioning}
\newcommand*{\hpchd}{\textsc{Basic}\xspace}
\newcommand*{\hpchdnewspapers}{\textsc{Extended}\xspace}
\newcommand*{\ecco}{\texttt{ECCO}\xspace}
\newcommand*{\eebo}{\texttt{EEBO-TCP}\xspace}
\newcommand*{\newspapers}{\texttt{Newspapers}\xspace}
\newcommand*{\rowstore}{\texttt{Aria}\xspace}
\newcommand*{\spark}{\texttt{Spark}\xspace}
\newcommand*{\columnstore}{\texttt{Columnstore}\xspace}
\newcommand*{\denorm}{\texttt{Denorm}\xspace}
\newcommand*{\intermediate}{\texttt{Intermediate}\xspace}
\newcommand*{\standard}{\texttt{Standard}\xspace}
\newcommand*{\edition}{\ensuremath{E}\xspace}
\newcommand*{\doc}{\ensuremath{D}\xspace}

\newcommand{\spara}[1]{\smallskip\noindent\textbf{#1}}
\newcommand{\nworks}{\texttt{n\_reuses}\xspace}
\newcommand{\sumnworks}{\texttt{sum\_n\_reuses}\xspace}

\newcommand*{\system}{\texttt{ReceptionReader}\xspace}

\newmintedfile[sqlcode]{sql}{
fontfamily=tt,
fontsize=\small,
linenos=true,
numbersep=5pt,
gobble=0,
mathescape,
}
\NewDocumentCommand{\sql}{v}{\texttt{#1}}

\usepackage{cleveref}

\begin{document}
\sloppy
\title{Optimizing a Data Science System for Text Reuse Analysis }

\author{Ananth Mahadevan, Michael Mathioudakis, Eetu M{\"a}kel{\"a} and Mikko Tolonen}
\affiliation{%
  \institution{University of Helsinki, Finland}
}




\begin{abstract}
\textit{Text reuse} is a methodological element of fundamental importance in humanities research: pieces of text that re-appear across different documents, verbatim or paraphrased, provide invaluable information about the historical spread and evolution of ideas, enable researchers to quantify the intellectual influence of specific authors and literary works, and sometimes offer precious clues in revealing the identity of an author or detecting plagiarism. Traditionally, humanities scholars have studied text reuse at a very small scale, for example, when comparing the opinions of two authors (e.g., 18th-century philosophers) as they appear in their texts (e.g., their articles). Against that backdrop, large modern digitized corpora offer the opportunity to revolutionize humanities research, as they enable the joint analysis of text collections that span entire centuries and the detection of large-scale patterns, impossible to detect with traditional small-scale analysis. For this opportunity to materialize, it is necessary to develop efficient data science systems that perform the corresponding analysis tasks. 

In this paper, we share insights from \system, a system for analyzing text reuse in large historical corpora. In particular, we describe the data management considerations that have allowed us to move from an originally \textit{deployed} but unoptimized version of the system to an extensively \textit{evaluated} and substantially optimized version, which is destined to be the newly deployed version of the system. The system is built upon large digitized corpora of 18th-century texts, which include almost all known books, articles, and newspapers from that period. It implements a data management and processing pipeline for billions of instances of text reuse. Its main functionality is to perform downstream analysis tasks related to text reuse, such as finding the instances of text reuse that stem from a given article or identifying the most reused quotes from a set of documents, with each task expressed as a database query. For the purposes of the paper, we discuss the related design options for data management, including various levels of database normalization and query execution, including combinations of distributed data processing (Apache Spark), indexed row store engine (MariaDB Aria), and compressed column store engine (MariaDB Columnstore). Moreover, we present an extensive evaluation in terms of various metrics of interest (latency, storage size, and computing costs) for varying workloads, and we offer insights from the trade-offs we observed and the choices that emerged as optimal in our setting. In summary, our results demonstrate that (1) for the workloads that are most relevant to text-reuse analysis, the row store engine (MariaDB Aria) emerges as the overall optimal choice for executing analysis tasks at the user-end of the pipeline, (2) big data processing (Apache Spark) is irreplaceable for all processing stages of the system's pipeline.

\end{abstract}

\maketitle

\pagestyle{\vldbpagestyle}


\section{Introduction}
Text reuse is an essential methodological element in humanities research, which involves the detection and analysis of similar pieces of text across different documents.
It offers a window into the historical evolution and dissemination of ideas, allowing researchers to trace intellectual influences, quantify the impact of specific authors and literary works, and occasionally unravel author identities or detect plagiarism instances. 
Traditionally, text reuse analysis has been conducted on a small scale, such as comparing the writings of two philosophers.
However, the advent of large-scale digitized historical corpora spanning centuries has transformed this field.
Modern algorithms enable the automatic detection of text reuses across documents, and modern technology enables the analysis of these reuses, uncovering large-scale reuse patterns that were previously indiscernible with smaller-scale analysis. 

In this paper, we present insights learned from \system, a system designed for analyzing text reuse in large historical corpora.
The \system system is built upon billions of pre-processed text reuses identified from 18th-century documents.
These documents include a vast array of books, articles, and newspapers from the 17th and 18th centuries.
They are primarily digitized using Optical Character Recognition (OCR) which tends to be noisy and error-prone.
The raw text reuses are obtained using a fuzzy-string-matching algorithm adapted to identify similar pieces of text between documents amidst noisy OCR data.
These reuses are then pre-processed by merging and clustering the individual text pieces.
This pre-processing pipeline results from continuous domain-driven research in digital humanities and is described in \Cref{sec:background}.
The system uses the pre-processed text reuses to provide functionality for two main downstream analysis tasks.
The first task, \textbf{reception}, is finding all the text reuses that stem from a particular document
The second task, \textbf{top quotes}, is identifying the most reused quotes from a set of documents.

We are interested in optimizing system performance metrics, such as query latency, data storage size, and computing costs for each downstream analysis task.
Hence, based on these metrics, we evaluate various design choices for the data management, storage and processing of the pre-processed text reuses. 
In particular, we consider three data management choices corresponding to three levels of database normalization: fully normalized data (referred to as \standard), task-agnostic normalized tables (\intermediate), and task-specific denormalized data (\denorm).
Moreover, we consider three frameworks for data storage and query execution: distributed data processing (Apache \spark), indexed row store database (MariaDB \rowstore), and compressed column store database (MariaDB \columnstore).
%
%
%


%
We extensively evaluate each design choice over various workloads for each downstream analysis task.
We study the trade-offs provided and discuss the insights that enabled the evolution from an initial, deployed, but unoptimized version to a highly optimized system for analyzing text reuses.  
From our results, we observe that the data management choices offer a trade-off between data storage size and query latency.
Specifically, materializing \denorm level data has the best query latency with increased data storage size, especially for the reception task.
On the other hand, the frameworks evaluated offer an independent trade-off between data storage, query execution costs and workload consistency.
For example, the \columnstore framework has a smaller data storage size compared to \rowstore for the same data normalization level.
However, the \columnstore framework has higher query latency resulting in larger cost for query execution in comparison to \rowstore. 
Similarly, the \spark framework has the highest query execution cost but is the only framework that completes task queries with a high workload.   

In summary, from our evaluation we find that the \rowstore framework, especially with \denorm level data, is optimal considering workloads relevant to the analysis of text reuses.
Furthermore, the \spark framework is irreplaceable for the pre-processing and materialization phases in the system.
Finally, the \columnstore framework is a good trade-off between data storage cost and query latency, however, is only feasible with the \denorm level normalized data.

The paper is organized as follows.
\Cref{sec:background} describes the text corpora, reuse detection method and the pre-processing pipeline of the \system system.
\Cref{sec:setup} discusses the datasets, metrics and design choices evaluated.
The queries and workloads for the reception and top-quotes tasks are described in \Cref{sec:reception-task,sec:top-quotes-task} respectively.  
In \Cref{sec:results}, we present the evaluation results for each design choice and discuss the trade-offs offered and key insights.
We highlight related works in \Cref{sec:related-works} and present our conclusions in \Cref{sec:conclusion}.
\section{Background}
\label{sec:background}
This section provides the necessary background to understand the \system system.
First, \Cref{sec:collections} introduces the three collections that make up our large historical text corpora.
Next, \Cref{sec:blast} discusses the algorithm used to detect text reuses across documents in the corpora.
Finally, \Cref{sec:pre-processing-pipeline} describes the phases in the pipeline for pre-processing the text reuses.

\subsection{Historical Text Corpora}
\label{sec:collections}

The historical text corpora used in our system come from different document collections shown in \Cref{tab:collections}. 
The Eighteenth Century Collections Online (\ecco) \cite{gale2003Ecco,Gregg2021Ecco} consists of English language books published in the 18th century digitized using a mix of OCR \cite{christy2018MassDigitizationEarly} and manual transcription \cite{galeEccoTcp}.
The Early English Books Online Text Creation Partnership (\eebo) \cite{mich2009Eebo} consists of accurately transcribed books spanning the 17th and 18th centuries.
The British Library Newspapers (\newspapers) \cite{galeBLNewspapers} collection contains over 300 years of published newspaper articles digitized using OCR.

There are two challenges in working with several document collections.
First, each collection has its own unique metadata format and different set of metadata attributes.
Such heterogeneous sources of metadata make downstream queries complex requiring several table joins and brittle to the addition of more collections.
Therefore, we collect, parse and standardize the metadata attributes required for downstream tasks such as publication dates, document titles and authors.
Second, research has shown that text from OCR is noisy and error-prone due to several factors, such as archaic fonts~\cite{gupta-2015-thesis}, document layouts~\cite{christy2018MassDigitizationEarly} and scanning quality~\cite{ye2013DocumentImageQuality}.
Furthermore, the quality of the OCR varies significantly within and without collections \cite{hill2019QuantifyingImpactDirty}.
Such OCR errors and variance in OCR noise pose challenges in identifying similar pieces of text.
We address this challenge by using novel algorithms that are robust to OCR noise and errors.

\begin{table}[ht]
    \centering
    \caption{Collections in the historical text corpora. Document length is measured in number of characters.}
    \label{tab:collections}
    \begin{tabular}{cccc}
        \toprule
        Collection & Publication Years & \# Docs& Avg.\ Doc Length \\
        \midrule
        \ecco & 1505-1839 & 207613& 295641\\
        \eebo & 1473-1865 & 60327& 18148\\
        \newspapers & 1604-1804 & 1769266& 9519\\
        
        \bottomrule
    \end{tabular}
\end{table}

\subsection{Text Reuse Detection}
\label{sec:blast}

Standard string matching algorithms perform poorly in identifying similar pieces of text in the presence of OCR noise.
Therefore, we use BLAST (Basic Local Alignment Search Tool) \cite{blast1990}, a fuzzy-string-matching algorithm that is widely used for DNA sequence matching and has proved to be more effective than alternatives in finding text reuses with noisy OCR data \cite{vesanto-etal-2017-applying}.
We adapt the algorithm as seen in \citet{vesanto-thesis-2018} to find similar pieces of text in a large corpora.
First, we encode the text of each document in the corpora into the format used by BLAST.
Then, using these encoded texts, we run the BLAST algorithm comparing every document against the entire corpora.
BLAST returns \emph{hits}, which are the similar pairs of text between documents.
Each hit contains a pair of document IDs along with the starting and ending offsets within the respective document's encoded text.
These encoded offset values are decoded to obtain the starting and ending offsets in the original text of the document.
Therefore, we define a \textbf{piece} as a fragment of text identified by the tuple of a document ID, the starting and ending offsets.
In addition, we define a \textbf{text reuse} as an undirected edge between similar pieces.

\begin{figure*}
    \centering
    
    \newcommand\file[2]{%
    \begin{scope}[xshift=#1cm,yshift=#2cm]
      \draw[gray,thick] (0,0) -- (0,1.5) -- (0.7,1.5) -- (0.7,1.1) -- (1,1.1) -- (1,0) -- cycle;
      \draw[gray,thick] (0.7,1.5) -- (1,1.1);
      \foreach \y in {0.2,0.3,...,1.1}{
         \draw[thick] (0.2,\y) -- (0.8,\y);
         }
      \foreach \y in {1.1,1.2,...,1.4}{
         \draw[thick] (0.2,\y) -- (0.65,\y);
         }
    \end{scope}
}

\begin{tikzpicture}
    \tikzset{
      piece/.style 2 args={red,draw,thick,rectangle,anchor=south west,minimum width=#1,minimum height=#2},
      defrag piece/.style 2 args={blue,draw,thick,rectangle,anchor=south west,minimum width=#1,minimum height=#2},
      source piece/.style 2 args={orange,draw,very thick,rectangle,anchor=south west,minimum width=#1,minimum height=#2},
      destination piece/.style 2 args={green!50!black,draw,very thick,rectangle,anchor=south west,minimum width=#1,minimum height=#2},
      reuse/.style={red,-,very thick},
      defrag reuse/.style={blue,-,very thick},
      reception/.style={black,-{Latex[scale=1]},very thick},
      cluster/.style={violet,-,very thick}
    }
    \begin{scope}[local bounding box=scope1]
    \file{0}{0}
    \node at (0.5,1.65) {A};
    \node[piece={0.8cm}{0.4cm}] at (0.1,0.1) (tr11) {};
    \node[piece={0.54cm}{0.4cm}] at (0.1,1.1) (tr12) {};
    \file{1.5}{1.5}
    \node at (2,3.15) {B};
    \node[piece={0.8cm}{0.4cm}] at (1.55,1.55) (tr21) {};
    \node[piece={0.78cm}{0.5cm}] at (1.64,1.63) (tr22) {};
    \file{3}{0}
    \node at (3.5,1.65) {C};
    \node[piece={0.54cm}{0.4cm}] at (3.1,0.95) (tr31) {};
    \node[piece={0.8cm}{0.4cm}] at (3.1,0.1) (tr32) {};
    \file{1.5}{-1.5}
    \node at (2,0.15) {D};
    \node[piece={0.8cm}{0.4cm}] at (1.55,-1.45) (tr41) {};
    \node[piece={0.78cm}{0.5cm}] at (1.63,-1.36) (tr42) {};
    \draw[reuse] (tr12.east) -- (tr21.west);
    \draw[reuse] (tr11.east) -- (tr41.west);
    \draw[reuse] (tr42.east) -- (tr32.west);
    \draw[reuse] (tr22.east) -- (tr31.west);
    \node at (2,-1.68) {};
    \end{scope}

    \begin{scope}[local bounding box=scope2,shift={(scope1.base east)},xshift=3cm]
    \file{0}{0}
    \node at (0.5,1.65) {A};
    \node[defrag piece={0.8cm}{0.4cm}] at (0.1,0.1) (tr11) {};
    \node[defrag piece={0.54cm}{0.4cm}] at (0.1,1.1) (tr12) {};
    \file{1.5}{1.5}
    \node at (2,3.15) {B};
    \node[defrag piece={0.8cm}{0.54cm}] at (1.58,1.6) (tr21) {};
    \file{3}{0}
    \node at (3.5,1.65) {C};
    \node[defrag piece={0.54cm}{0.4cm}] at (3.1,0.95) (tr31) {};
    \node[defrag piece={0.8cm}{0.4cm}] at (3.1,0.1) (tr32) {};
    \file{1.5}{-1.5}
    \node at (2,0.15) {D};
    \node[defrag piece={0.8cm}{0.54cm}] at (1.6,-1.4) (tr41) {};
    \draw[defrag reuse] (tr12.east) -- (tr21.west);
    \draw[defrag reuse] (tr11.east) -- (tr41.west);
    \draw[defrag reuse] (tr41.east) -- (tr32.west);
    \draw[defrag reuse] (tr21.east) -- (tr31.west);
    
    \draw[cluster] plot [smooth cycle] coordinates{(0.1,1) (3.8,0.95) (3.5,2) (2,2.25) (0.6,2) };
    \draw[cluster] plot [smooth cycle] coordinates{(0.0,0.6)  (4,0.6) (3.5,-1) (2,-1.75) (0.55,-1)};
    \end{scope}

    \begin{scope}[local bounding box=scope3,shift={(scope2.base east)},xshift=3cm]
      \file{0}{0}
      \node at (0.5,1.65) {A};
      \node[source piece={0.8cm}{0.4cm}] at (0.1,0.1) (tr11) {};
      \node[destination piece={0.54cm}{0.4cm}] at (0.1,1.1) (tr12) {};
      \file{1.5}{1.5}
      \node at (2,3.15) {B};
      \node[source piece={0.8cm}{0.54cm}] at (1.58,1.6) (tr21) {};
      \file{3}{0}
      \node at (3.5,1.65) {C};
      \node[destination piece={0.54cm}{0.4cm}] at (3.1,0.95) (tr31) {};
      \node[destination piece={0.8cm}{0.4cm}] at (3.1,0.1) (tr32) {};
      \file{1.5}{-1.5}
      \node at (2,0.15) {D};
      \node[destination piece={0.8cm}{0.54cm}] at (1.6,-1.4) (tr41) {};
      \draw[reception] (tr11.east) -- (tr41.west);
      \draw[reception] (tr11.east) -- (tr32.west);

      \draw[reception] (tr21.west) -- (tr12.east);
      \draw[reception] (tr21.east) -- (tr31.west);
  
      \draw[cluster] plot [smooth cycle] coordinates{(0.1,1) (3.8,0.95) (3.5,2) (2,2.25) (0.6,2) };
      \draw[cluster] plot [smooth cycle] coordinates{(0.0,0.6)  (4,0.6) (3.5,-1) (2,-1.75) (0.55,-1)};
      \end{scope}

      \draw [->,black,very thick] ($(scope1.east)!0.1!(scope2.west)$) -- (scope2.west) node[midway,above]{Defragment} node[midway,below] {\& Cluster};
      \draw [->,black,very thick] (scope2.east) -- (scope3.west) node[midway,above]{Identify Sources};

    \matrix [draw,matrix anchor = south,yshift=5mm] at ($(scope1.north west)!0.5!(scope3.north east)$) {
    \node [inner sep=0,label=right:Doc,anchor=south west] at (0,-0.5mm){\faFileTextO}; & 
    \node [piece={0.5cm}{0.1cm},label=right:Piece] {}; & 
    \node[reuse,inner sep=0,minimum width=4mm,draw,anchor=south west,label=right:Text Reuse] at ++(0,1mm) {}; & 
    \node [defrag piece={0.5cm}{0.1cm},label=right:Defrag Piece] {}; & 
    \node[defrag reuse,inner sep=0,minimum width=4mm,draw,anchor= south west,label=right:Defrag Text Reuse] at ++(0,1mm) {}; & 
    \node[cluster,inner sep=0,minimum width=4mm,draw,anchor= south west,label=right:Cluster] at ++(0,1mm) {};& 
    \node [source piece={0.5cm}{0.1cm},label=right:Source Piece] {}; & 
    \node [destination piece={0.5cm}{0.1cm},label=right:Destination Piece] {}; & 
    \draw[reception] (0,1mm) -- (5mm,1mm) node[inner sep =0,minimum size =0,right,anchor=west] {Reception};
    \\
    };
\end{tikzpicture}

%

    \caption{Illustration of the pre-processing pipeline for text reuses. The hits from BLAST (red) are first defragmented (blue) and then clustered (purple). Then in each cluster, the piece from the earliest document is identified as the source (yellow) and the rest are destinations (green). The directed edge between a source and destination piece is a reception edge. }
    \label{fig:pre-processing-pipeline}
\end{figure*}
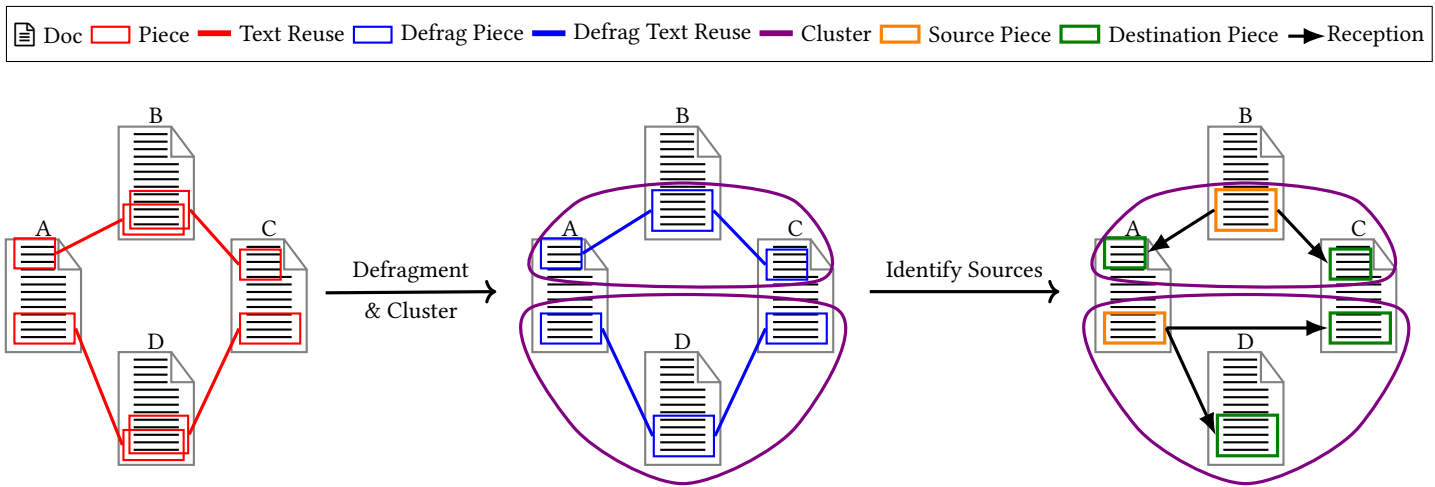

\subsection{Pre-processing Pipeline}
\label{sec:pre-processing-pipeline}

The text reuses obtained as hits using the BLAST algorithm require pre-processing before they are useful for downstream tasks.
This pipeline is illustrated in \Cref{fig:pre-processing-pipeline} and its three phases are as follows.

\spara{Phase 1: Defragmentation.}
When a paragraph of text is reused in several documents, it's starting and ending offsets in different BLAST hits might differ slightly due to the OCR noise and the fuzzy-string matching.
This phenomenon is known as \emph{fragmentation} and is depicted as overlapping red boxes in documents B and D in \Cref{fig:pre-processing-pipeline}.
Such fragmentation leads to issues in the downstream tasks, either due to redundancy (multiple overlapping pieces referring to the same text) or loss of results (e.g., identifying only short matches rather than longer matching passages).
%
%
To remedy this, we perform a ``conservative'' merging of overlapping pieces within a document.
This only merges pieces of similar length that overlap a given amount, thereby ensuring that smaller pieces do not get merged into larger overlapping pieces.
This behavior is beneficial because smaller pieces have a different contextual meaning than larger text reuse pieces.
The conservative merging process results in new \emph{defragmented pieces} and the mapping between defragmented pieces are the new \emph{defragmented text reuses} depicted by blue boxes and lines in \Cref{fig:pre-processing-pipeline} respectively.

\spara{Phase 2: Clustering.}
We create a reuse network where defragmented pieces are nodes and text reuse pairs are edges.
This network is clustered to find all the similar text pieces across documents.
For the network clustering we use Chinese Whispers \cite{biemann-2006-chinese}, a label-propagation graph clustering algorithm.
The resulting clusters are groups similar of pieces from different documents depicted by purple enclosures in \Cref{fig:pre-processing-pipeline}.

\spara{Phase 3: Source identification.}
In each cluster, we identify the earliest piece(s) using the publication dates from the document metadata.
These earliest pieces are called the ``sources'' of the text reuse in that cluster.
Furthermore, the pieces which are not sources in a cluster are called the ``destination'' of the text reuse in that cluster.
Hence, an edge from a source piece to a destination piece signifies the direction of the text reuse and is called a \emph{reception edge}.
Due to the granularity of the publication date metadata from different collections, there can be several source pieces in each cluster.
In such scenarios, we treat each source piece as equal and possible originations for the text reuse in a cluster. 
Hence, in such multi-source clusters, each destination piece will have multiple incoming reception edges.
The reception edges are used to find text reuses that either originate from or end on particular documents for downstream tasks. 

\begin{figure}[t]
    \centering
    \includegraphics[width=0.45\textwidth]{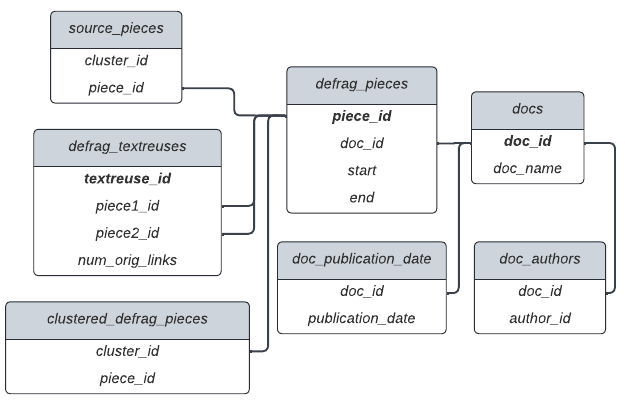}
    \caption{Simplified data schema for pre-processed text reuses}
    \label{fig:erd}
\end{figure}

For the purposes of this paper, the three-phase pre-processing pipeline described above is considered \textit{fixed}, as it has been developed through years of field expertise by the digital humanities researchers in our team.
The pre-processing pipeline has been implemented in Apache Spark and can scale to billions of raw BLAST hits.
At the end of the pre-processing pipeline, we obtain data in the form of several Spark data frames stored as Parquet files.
We omit the nuances of the complex document-metadata relationships and present a simplified schema shown in \Cref{fig:erd}.
In the following sections, we will use this simplified schema to explain the \system system and the downstream tasks.

\section{Evaluation Setup}
\label{sec:setup}

In this section, we describe the setup of the experimental evaluation.
First, \Cref{sec:datasets} presents the two text reuse datasets used to evaluate the system.
Then, in \Cref{sec:normalization-levels}, we discuss the three data management design choices expressed as levels of database normalization.
Next, \Cref{sec:data-storage-and-query-execution} presents the three frameworks for data storage and query execution.
Additionally, \Cref{sec:processing-and-loading} describes how to materialize and load the data for each normalization level and framework combination. 
Finally, \Cref{sec:implementaion-details,sec:metrics} discuss the implementation details of the frameworks and the system evaluation metrics respectively.

\subsection{Datasets}
\label{sec:datasets}
\begin{table}[ht]
    \centering
    \caption{Text Reuse Datasets}
    \label{tab:datasets}
    \begin{tabular}{ccc}
    \toprule
    \multirow{2}{*}{Attribute} & \multicolumn{2}{c}{Dataset} \\\cmidrule(l){2-3}
                           & \hpchd        & \hpchdnewspapers \\\cmidrule(l){1-3}
    \multirow{2}{*}{Collections} & \ecco \&  & \ecco, \eebo \\
                                & \eebo & \& \newspapers \\
    \# documents &  267940& 2037206\\ 
    BLAST hits & 965665512 & 6310546114 \\
    \# defragmented pieces & 383894269 &  1098916260\\
    Avg.\ defragmented piece length & 748 & 461 \\
    \# defragmented reuses & 965134100 & 5608256233 \\
    \# clusters &  50332146& 91576355\\ 
    \# authors &  46017& 46017\\ 
    \bottomrule
    \end{tabular}
\end{table}

Following the steps mentioned in \Cref{sec:blast,sec:pre-processing-pipeline} we collect two datasets of text reuse, shown in \Cref{tab:datasets}.
The dataset used in our originally deployed \system system is called \hpchd and consists of the text reuses between documents, primarily books, from the \ecco and \eebo collections. 
The dataset starts with nearly one billion hits from the BLAST algorithm for fuzzy string matching. 
After pre-processing there are $383$ million pieces in $47$ million text reuse clusters.

The second dataset, \hpchdnewspapers, consists of text reuses between documents from all the collections in \Cref{tab:collections}.
This dataset contains nearly $6.5\times$ BLAST hits compared to the \hpchd dataset due to the inclusion of the \newspapers collection.
We see similar increases after pre-processing resulting in $2.8\times$ more defragmented pieces and $1.9\times$ more text reuse clusters. 
However, note that the number of authors do not increase because the \newspapers collection's metadata does not contain authorship attributes.   
Furthermore, newspaper articles are smaller than books, resulting in an average piece being $1.6\times$ shorter.
We use the \hpchdnewspapers dataset in our evaluation to showcase the scalability of the \system system.

\subsection{Data Normalization Levels}
\label{sec:normalization-levels}
We consider three design choices for data management expressed as three levels of database normalization.
Each level of normalization consists of different materialized data tables for each downstream analysis task.
The levels range from using fully normalized data to task-specific denormalized data.
On one end, fully normalized data reduces redundancies and used lesser storage space while making queries for downstream tasks more computationally demanding.
On the other end, materializing task-specific data requires more storage space while making the queries for downstream tasks faster.
We now describe the three normalization levels.

First, the \standard normalization level uses fully normalized data as per the schema shown in \Cref{fig:erd}.
The queries for downstream tasks will require joining several tables and filtering them to obtain the desired result. 

Second, the \intermediate normalization level materializes data which is task-agnostic and can be used in each of the downstream tasks.
Specifically, we materialize the pieces in each text reuse cluster which are the destination of the reuses (i.e., not the source pieces).
We call this table \sql{destination_pieces} and the SQL query to materialize it is shown in \Cref{sql:non-source-pieces}.
The SQL query performs an exclusive right join on the \sql{source_pieces} and \sql{clustered_defrafg_pieces} tables from \Cref{fig:erd} to obtain the pieces and their respective clusters which are not in the \sql{source_pieces} table.
These destination pieces are task-agnostic and useful in both the reception and top quotes tasks.
Therefore, the main difference between the queries for the \standard and \intermediate normalization levels is whether the \sql{destination_pieces} table is materialized.

\begin{listing}
    \sqlcode{non-source-pieces.sql}
    \caption{SQL for \texttt{destination\_pieces} }
    \label{sql:non-source-pieces}
\end{listing}

Third, the \denorm normalization level materializes task-specific data that aids in answering the queries of the specific downstream task.
For each downstream task, we first express the task-specific data as SQL queries which use the normalized tables seen in \Cref{fig:erd}.
Then, we materialize the data in the system and use it while answering the downstream queries.
We describe the \denorm level data and queries for each task in the upcoming sections.

\subsection{Data Storage and Query Execution}
\label{sec:data-storage-and-query-execution}
We evaluate the \system system on three frameworks for data storage and query execution. 
Each framework has a unique format to store the data and a corresponding SQL execution engine to answer downstream tasks expressed as SQL queries.
The different frameworks and their components are shown in \Cref{tab:storage-and-execution}.
\begin{table}[h]
    \centering
    \caption{Storage formats and query execution engines for different frameworks  }
    \label{tab:storage-and-execution}
    \begin{tabular}{ccc}
    \toprule
    Framework & Storage Format & Query Execution\\
    \midrule
     \spark & Apache Parquet & Apache Spark SQL\\
     \rowstore & indexed row-store & MariaDB Aria\\
     \columnstore& compressed column-store & MariaDB ColumnStore \\
    \bottomrule
    \end{tabular}
\end{table}

In the \spark framework, the data is stored as Parquet files on a cloud object storage platform.
These Parquet file objects are loaded in \spark as distributed data frames and then used to answer SQL queries.
In both the relational \rowstore and \columnstore frameworks, a MariaDB database is created, and the data are loaded into tables in their respective storage formats.
Additionally, for the \rowstore framework, we create B+-tree indexes on the essential columns of each table.
The downstream tasks, expressed as SQL queries, are run on the framework's execution engines with the appropriate tables to obtain results.

\subsection{Data Materialization and Loading}
\label{sec:processing-and-loading}
At the end of the pre-processing pipeline described in \Cref{sec:pre-processing-pipeline}, the text reuse data is stored as Parquet files in the normalized schema shown in \Cref{fig:erd}, which corresponds to the \standard normalization level (as explained in~\Cref{sec:normalization-levels} above).
Obtaining the \intermediate and \denorm from the \standard normalization level requires materializing additional data.
%
%
%
Below we describe two approaches for the data materialization, depending on whether we use \spark or the relational frameworks (\rowstore or \columnstore) to perform it.

The first approach is to materialize all the necessary data using the \spark framework and then bulk-load the data into the \rowstore and \columnstore frameworks.
Towards this, we execute the materialization SQL query for a given normalization level in the \spark framework and store the data as Parquet files.
Then, we load the necessary Parquet files into the \rowstore and \columnstore frameworks as tables.
%
%
%

The second approach is to use the  relational frameworks (\rowstore and \columnstore) frameworks to perform the materialization directly in their storage formats.
Towards this, we first load all the pre-processed data into the \rowstore and \columnstore frameworks.
Then, in each framework, we execute the materialization SQL query for a given normalization level and store the result as a table in the respective storage format.

\subsection{Framework Implementation}
\label{sec:implementaion-details}

We use the cloud computing resources provided by CSC\footnote{\url{https://research.csc.fi/cloud-computing}} to implement and evaluate the different frameworks described in \Cref{sec:data-storage-and-query-execution}.

The \spark framework is implemented as a Kubernetes cluster on CSC's Rahti service.
The cluster has 38 worker pods and one driver pod.
Each worker (and driver) pod has two cores and 32 GiB RAM.
Only the driver pod has 16 GiB persistent storage for storing code and runtime environments.
All the data are stored in cloud data buckets on CSC's Allas service and accessed in \spark using the S3 API\footnote{\url{https://spark.apache.org/docs/latest/cloud-integration.html}}.

The \rowstore and \columnstore frameworks are implemented on CSC's Pouta service.
We create a Virtual Machine (VM) which has 16 cores, 78 GiB RAM and 700 GiB SSD storage. 
Then we attach a persistent cloud storage volume to the VM and create ZFS pool and designate the SSD as cache. 
This configuration offers larger storage space as well as better I/O performance.

There are resource costs for running the system in CSC measured in Billing Units (BU).
In \Cref{tab:storage-execution-costs}, we report data storage and query execution costs for each framework as per our implementation.  

\begin{table}
    \centering
    \caption{Data storage and query execution costs of different frameworks in CSC Billing Units (BU)}
    \label{tab:storage-execution-costs}
    \begin{tabular}{ccc}
        \toprule
        Framework & Data Storage Cost & Query Execution Cost \\
        \midrule
        \spark & 1 BU/TiB~hr & 1254 BU/hr\\
        \rowstore & 3.5 BU/TiB~hr & 24 BU/hr\\
        \columnstore & 3.5 BU/TiB~hr & 24 BU/hr\\
        \bottomrule
    \end{tabular}
\end{table}

\subsection{Metrics of Interest}
\label{sec:metrics}

Each downstream analysis task is expressed as a database query and executed on the framework of the system.
Hence, the system performance for a given query is affected by the design choices as well as the downstream task.
We evaluate the performance of the system using the three important metrics described below.

\spara{Query Latency} 
The time duration to obtain a result set for a query is defined as the query latency.
In practice, we measure latency as the time taken to iterate through all the rows from the result set of a query.
Furthermore, to mitigate the effect of network latency on the query latency, in the \rowstore and \columnstore frameworks, we execute the queries directly on the VM where the data is stored.
A low query latency is desirable as it leads to better user experience.
Furthermore, the workload of a query determines the computational effort required to return a result set which directly affects the query latency.
For our experiments, we define the workload of each downstream analysis query and evaluate the system over different query workloads.  

\spara{Data Storage Size and Cost}
We define the data storage size as the total size of the materialized data required to answer a given query.
In practice, we measure size of the materialized data as the disk space used.
By its definition, the data storage size depends on the normalization level and the framework of the system.
The costs for data storage in different frameworks is shown in \Cref{tab:storage-execution-costs} measured in CSC Billing Units per tebibyte hour (BU/TiB hr). 
A low query storage size is desirable as it leads to lower costs for storing the data and indicates better scalability.
In our experiments, for a given framework, we calculate the data storage size in TiB for each combination of normalization level and downstream task and report the total storage cost in BU/hr.

\spara{Query Execution Cost}
The cost of computational resources required to answer a downstream query is defined as the query execution cost.
In practice, we use the execution cost for a given framework shown in \Cref{tab:storage-execution-costs}, and the query latency to compute the cost to execute one query in that framework.
A low query execution cost is desirable as it reduces the cost of running the system and in turn allows scaling to multiple users.
%

\section{Reception Task}
\label{sec:reception-task}

In this downstream analysis task, the aim is to find all the text reuses that originate from a given a query document, denoted by \doc.
\Cref{sec:reception-queries} describes the queries to achieve this task at each level of data normalization.
Then, \Cref{sec:reception-workload-distribution} quantifies the workload of a reception query and analyzes the distribution of the workload.
Lastly, we describe the process of sampling the workload distribution to get query documents for the system evaluation.  

\subsection{Reception Queries}
\label{sec:reception-queries}
The steps to find the text reuses that originate from a document \doc are as follows.
First, we find all the clusters where the source piece originates from \doc.
Next, we gather all the destination pieces of those clusters. 
Lastly, we create a mapping between a source piece and each destination piece in a cluster.
These mappings are the \emph{reception edges}, and we return them as the result set.

Following the above steps, the SQL query for the \intermediate normalization level is shown in \Cref{sql:reception-intermediate}.
This normalization level contains data from the schema shown in \Cref{fig:erd} and the additional materialized \sql{destination_pieces} table.
The SQL query requires one join to find the source clusters and a further two joins to find the text reuses. 
In the result set, each row contains one reception edge between a source piece from the given document \doc and a destination piece. 

\begin{listing}
    \sqlcode{reception-intermediate.sql}
    \caption{\intermediate level Reception query}
    \label{sql:reception-intermediate}
\end{listing}

The only difference for the \standard normalization level is that the \sql{destination_pieces} data is not materialized.
Therefore, the \standard level reception query contains the additional subquery shown in \Cref{sql:non-source-pieces}. 

For the \denorm normalization level, we materialize a table called \sql{reception_edges_denorm}, containing the reception edges for all the documents in the dataset.
The attributes of the table and the SQL query for materializing it are shown in \Cref{tab:reception-edges-denorm-table} and \Cref{sql:reception-edges-denorm} respectively.
Once the table is materialized, the \denorm level reception query is a simple filtering query as seen in \Cref{lst:reception-denorm}.

The materialization query shown in \Cref{sql:reception-edges-denorm} first creates a mapping from each source piece ID in a cluster to every destination piece ID of that cluster.
This mapping, called the \sql{reception_edges} subquery, is denormalized to obtain the document ID and offsets for the source and destination pieces.
\Cref{tab:reception-edges-denorm-stats} shows the statistics of the materialized table for different datasets. 
We see that each dataset has more than a billion reception edges.
Furthermore, the \hpchdnewspapers dataset has $3.7\times$ more reception edges than the \hpchd dataset. 
This large number of reception edges will affect the performance of the \denorm level query in different frameworks.

\begin{table}
    \centering
    \caption{Reception \denorm table attributes}
    \label{tab:reception-edges-denorm-table}
    \begin{tabular}{cc}
        \toprule
        \multicolumn{2}{c}{\texttt{reception\_edges\_denorm}}\\
        \midrule
        Attribute & Description\\
        \midrule
        \sql{src_doc_id}&Source piece document ID\\
        \sql{src_start}& Source piece start offset\\
        \sql{src_end}&Source piece end offset\\
        \sql{dst_doc_id}&Destination piece  document ID\\
        \sql{dst_start}&Destination piece start offset\\
        \sql{dst_end}&Destination piece end offset\\
        \bottomrule
    \end{tabular}
\end{table}

\begin{listing}
    \sqlcode{reception-edges-denrom.sql}
    \caption{Reception task \denorm level materialization query}
    \label{sql:reception-edges-denorm}
\end{listing}

\begin{listing}
    \sqlcode{reception-denrom.sql}
    \caption{\denorm level Reception query }
    \label{lst:reception-denorm}
\end{listing}

\begin{table}
    \centering
    \caption{Statistics of \texttt{reception\_edges\_denorm} table for different datasets}
    \label{tab:reception-edges-denorm-stats}
    \begin{tabular}{crr}
        \toprule
        Statistic & \hpchd & \hpchdnewspapers \\
        \midrule
        \# of reception edges   & 1277583845 & 4739160201\\
        \# of source documents & 227917 & 1384487 \\
        \# of destination documents & 246437  & 1735188 \\
        \bottomrule
    \end{tabular}
\end{table}

\subsection{Reception Workload Distribution}
\label{sec:reception-workload-distribution}
\begin{figure}
    \centering
    {\includegraphics[width=\linewidth]{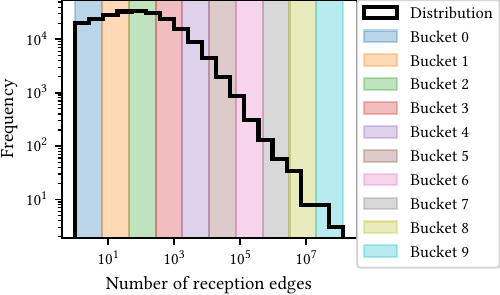}}
    \caption{Reception task workload distribution for the \hpchd dataset. The log-spaced buckets, shown as vertical bands, are used for sampling query documents for system evaluation.}
    \label{fig:reception-workload-distribution}
\end{figure}

The workload of a reception query is quantified by the number of reception edges for the given document \doc.
\Cref{fig:reception-workload-distribution} shows the distribution of the number of reception edges for the \hpchd dataset.
%
%
We see that on one end there are a lot of documents with only one reception edge and on the other end there are a few documents with more than 100 million text reuses.
This indicates that the workload follows a power-law distribution with a heavy-tail.
We see a similar distribution for the larger \hpchdnewspapers dataset.

To effectively evaluate the impact of the workload on the system performance, we sample $10$ documents from $10$ logarithmic spaced buckets labelled 0 to 9 shown in \Cref{fig:reception-workload-distribution}.
Documents from higher numbered buckets have more reception edges causing reception queries to take longer and vice versa. 
These $100$ sampled documents of varying workloads are used to study the trade-offs offered by different normalization levels and frameworks.

\section{Top Quotes Task}
\label{sec:top-quotes-task}
In this task, we are given a set of documents called an \emph{edition} denoted by \edition.
The goal is to find the top 100 most reused quotes from those documents.
Specifically, based on domain expertise, we define a quote as a piece of text between 150 and 300 characters long that has been reused in documents written by a different author at least once.
In \Cref{sec:top-quotes-queries}, we describe the queries for the task at different data normalization levels. 
Then, in \Cref{sec:top-quotes-workload-distribution}, we quantify and study the workload of a top quotes query.
Finally, we describe the process to sample editions from the query workload distribution to evaluate the system.

\subsection{Top Quotes Queries}
\label{sec:top-quotes-queries}
The importance of a quote is quantified by the number of times it is reused in documents written by a different author, denoted by \nworks.
Given an edition set \edition, the three steps for computing the \nworks statistic and finding the top 100 quotes are as follows.
%
%

%
First, we find all clusters where a source piece originates from the documents in \edition and is between 150 and 300 characters long.
%
%
Second, we find the authors of the documents in \edition.
%
%
Third, for each source piece cluster from step one, we gather the destination pieces and find the document and the author of each destination piece.
Then, using the authors of \edition from the second step, we count the number of unique destination documents with different authors.
This count is the \nworks statistic for the cluster, which is mapped back to the source pieces of that cluster.
Finally, we sort the source pieces in descending order of the \nworks statistic and return the first 100 pieces as the top quotes.

\begin{listing}[h]
    \sqlcode{quotes-intermediate.sql}
    \caption{\intermediate level Top Quotes query}
    \label{sql:quotes-intermediate}
\end{listing}
The \intermediate level SQL query following the above steps is shown in \Cref{sql:quotes-intermediate}.
The first two steps are simple subqueries performing table joins and filtering based on the query set \edition and piece length.
The third step is a group-by aggregation over clusters using the materialized \sql{destination_pieces} table. 
The final result set is obtained after ordering the source pieces based on \nworks and limiting them to the top 100 rows.

The only change for the \standard level SQL query is the inclusion of the subquery for the \sql{destination_pieces} table shown in \Cref{sql:non-source-pieces}.

For the \denorm level, we materialize a table with the \sql{n_works} statistic pre-computed for each source piece in the dataset regardless of length.
The materialization SQL query is similar to the query shown in \Cref{sql:quotes-intermediate} without the first filtering subquery.
We refer to this materialized table as \sql{source_piece_statistics_denorm} and its attributes are described in \Cref{tab:source-piece-statistics-denorm-table}. 
The \denorm level SQL query then filters the materialized table based on the documents in edition \edition and piece length as shown in \Cref{sql:quotes-denorm}. 
The statistics of the materialized table for different datasets is shown in \Cref{tab:source-piece-statistics-denorm-stats}.
The \hpchdnewspapers dataset has $1.6\times$ more source pieces from $6\times$ from documents.
Furthermore, the number of editions in both datasets are similar to the number of documents.
Therefore, each edition set contains a few documents.
%
%
%
%
%
%

\begin{table}
    \centering
    \caption{Top Quotes \denorm table attributes}
    \label{tab:source-piece-statistics-denorm-table}
    \begin{tabular}{cc}
        \toprule
        \multicolumn{2}{c}{\texttt{source\_piece\_statistics\_denrom}} \\
        \midrule
        Attribute & Description\\
        \midrule
        \sql{piece_id}&Source piece id\\
        \sql{cluster_id}& Cluster id\\
        \sql{doc_id}& Source document id\\
        \sql{start}& Piece start offset\\
        \sql{end}&Piece end offset\\
        \sql{piece_length}& Length of piece\\
        \sql{n_reuses}& \# reuses from other authors\\
        \bottomrule
    \end{tabular}
\end{table}
\begin{table}
    \centering
    \caption{Statistics of \texttt{source\_piece\_statistics\_denorm} table for different datasets}
    \label{tab:source-piece-statistics-denorm-stats}
    \begin{tabular}{crr}
        \toprule
        Statistic & \hpchd & \hpchdnewspapers \\
        \midrule
        \# source pieces  & 56093986 & 90144871\\
        \# clusters & 48314577 & 77733584\\
        \# documents & 227917 & 1384487 \\
        \# editions & 209305 & 1365831 \\
        Avg.\ piece length & 1236  & 964 \\
        Avg.\ \nworks & 2.6966  & 6.6588 \\
        \bottomrule
    \end{tabular}
\end{table}
\begin{listing}
    \sqlcode{quotes-denorm.sql}
    \caption{The \denorm query for top quotes task}
    \label{sql:quotes-denorm}
\end{listing}

\subsection{Top Quotes Workload Distribution}
\label{sec:top-quotes-workload-distribution}
We now quantify the workload of a top quotes query for a given edition set \edition.
No additional work is required for the \denorm level because the \nworks statistic is pre-computed.
However, for the \standard and \intermediate level queries, computing the \nworks statistic is the main workload, which depends on two factors.
The first factor is the number of source pieces from the documents in \edition.
The second factor is the number of destination pieces for each source piece.
This second factor correlates with the \nworks statistic.
Furthermore, the sum of the \nworks statistic for a given edition set, denoted by \sumnworks, correlates with the overall computational workload.
Therefore, we use the \sumnworks statistic to quantify the workload of a top quotes query.

\begin{figure}
    \centering
    {\includegraphics[width=\linewidth]{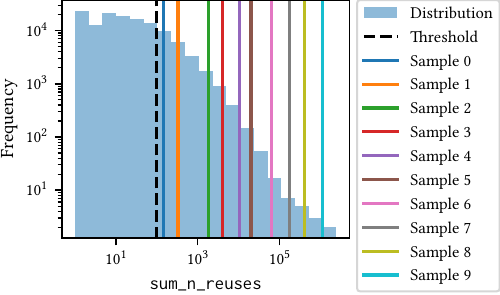}}
    \caption{Top Quotes task workload distribution (\sumnworks) for the \hpchd dataset. The threshold for sampling at $100$ is shown as a vertical dashed line.  The workload of the $10$ sampled edition sets are shown as solid vertical lines. }
    \label{fig:quotes-workload}
\end{figure}

\Cref{fig:quotes-workload} displays the distribution of the \sumnworks statistic for all edition sets in the \hpchd dataset. 
Similar to the reception task, the workload distribution follows a power-law curve with a few edition sets having greater than a million other documents reusing the text from the edition.
However, we also see that some edition sets have \sumnworks as low as one.
Therefore, to effectively sample from the workload distribution, we set a lower threshold of $100$ for the \sumnworks statistic.
This threshold ensures that, on average, each source piece in the top 100 quotes of a sampled edition will have at least one other author's document reusing it. 
From the thresholded \sumnworks distribution, we sample $10$ editions, each from one of $10$ log-spaced buckets numbered from $0$ to $9$.
These \sumnworks values for the sampled edition are shown \Cref{fig:quotes-workload} as colored vertical lines.
Similarly, $10$ editions are sampled from the workload distribution of the \hpchdnewspapers dataset.
These sampled editions are used as the query edition \edition to evaluate the system performance for different design choices.

\section{Results and Discussion}
\label{sec:results}

%
As discussed in \Cref{sec:setup}, our evaluation is structured around the two core design choices of database normalization levels and frameworks for data storage \& query execution.   
Each design choice has three options namely \standard, \intermediate and \denorm normalization levels and the \spark, \rowstore and \columnstore frameworks.
%
%
%
We evaluate each design choice combination on metrics such as query latency, data storage size, and computing costs.
The aim of this evaluation is to identify the most effective combination of design choices for optimizing the system's performance under various workloads for the two downstream text reuse analysis tasks.


For each combination of normalization level and framework the evaluation is conducted as follows.
First, we gather the text reuse data for the given dataset from the pre-processing pipeline.
Then, we materialize any additional data using the \spark framework and, if necessary, load the data into the framework being evaluated.
Next, for each document sampled from the reception workload distribution (see~\Cref{sec:reception-workload-distribution}), we execute the reception query with a timeout of five minutes.
Similarly, for each edition set sampled from the workload distribution of the top quotes task (see~\Cref{sec:top-quotes-workload-distribution}), we execute the top quotes query with a timeout of 15 minutes.  
For each query executed, we collect the query latency and other metrics.

The results from the evaluation are reported and discussed in the following sections.
\Cref{sec:post-processing-and-loading-results} discusses the practicalities in materializing data for the \intermediate and \denorm levels and loading them into the \rowstore and \columnstore frameworks. 
Then, \Cref{sec:workload-results} presents the query latency results across different workloads for each downstream tasks and discusses the impact of each design choice.
Lastly, \Cref{sec:cost-results} presents the trade-offs between data storage and query execution costs for each design choice. 




\subsection{Materialization and Loading Results}
\label{sec:post-processing-and-loading-results}
First, we discuss the practicalities related to materializing the data for the \intermediate and \denorm levels and then the loading the data into the \rowstore and \columnstore frameworks. 
Then, we present the materialization and loading time results and discuss the insights.

\spara{Materialization Practicalities.}
\Cref{sec:processing-and-loading} describes two approaches for materializing the data required for the \intermediate and \denorm levels.
The first approach uses the \spark framework to execute SQL queries and materialize the data as Parquet files.
The second approach loads the \standard data into the relational frameworks (\rowstore and \columnstore) and uses the relational engine to execute SQL queries to materialize the data as tables. 
However, in our experiments, we found that the second approach is only possible when using the \rowstore framework and the \intermediate level.
Even in this case, the time taken to materialize the table and index it in \rowstore was twice as long as the first approach.
Furthermore, for the \denorm level, the second approach failed to complete after 24 hours in the \rowstore framework and reported a memory error in the \columnstore framework.
Therefore, in this section, we only report the results from using the \spark framework for materialization.

\spara{Data Loading Practicalities.}
After the data is materialized using \spark, it is loaded into the \rowstore and \columnstore frameworks.
For the \rowstore framework, we turn off database transactions for tables and load data in parallel from \spark using the JDBC API\footnote{\url{https://spark.apache.org/docs/latest/sql-data-sources-jdbc.html}}.
However, database transactions cannot be disabled for tables in the \columnstore framework, resulting in infeasible loading times for large tables using JDBC.
Therefore, we use the \texttt{cimport}\footnote{\url{https://mariadb.com/docs/server/data-operations/data-import/enterprise-columnstore/cpimport/}} tool to instead bulk-load data from tables in the \rowstore framework into \columnstore.
Furthermore, because loading sorted tables into \columnstore increases the system performance, we first index the \rowstore tables and then load sorted data into \columnstore.
%

%
\Cref{tab:materialization-results} reports the time taken to then load the data for each normalization level into \rowstore and \columnstore after it has been materialized using \spark for the \hpchdnewspapers dataset. 
For the \rowstore framework, the reported loading time includes the time spent indexing the table.
Similarly, due to the practicalities mentioned above, the loading time reported for the \columnstore framework consists of the time spent in loading the data into and indexing the tables in \rowstore along with the time taken to load into \columnstore.

Let us first analyze the \intermediate level results from \Cref{tab:materialization-results}. 
Materializing the \sql{destination_pieces} data in \spark only takes a few minutes.
Then, bulk-loading the data into \rowstore takes $24$ minutes and then indexing the table takes $51$ minutes.
It then takes $16$ more minutes to bulk-load the indexed data from \rowstore to \columnstore.
    
\begin{table}
    \centering
    \caption{Materialization and Loading times (in seconds) by normalization level and framework for the \hpchdnewspapers dataset. Each cell shows the loading time with the materialization time using \spark in parentheses.}
    \label{tab:materialization-results}
    \begin{tabular}{crr}
        \toprule
        \multirow{2}{*}{Normalization} & \multicolumn{2}{c}{Framework} \\\cmidrule(l){2-3}
                           Level &  \rowstore & \columnstore \\
        \midrule
        \intermediate & 4552 (122) & 5519 (122) \\
        \denorm(Reception Task) & 24550 (690) & 52222 (690) \\ 
        \denorm(Top Quotes Task) & 639 (1472) & 953 (1472) \\ 
        \bottomrule
    \end{tabular}
\end{table}

Next, we analyze the \denorm level results for the reception and top quotes from \Cref{tab:materialization-results}.
For the reception task, the materialization is fast in \spark taking $11$ minutes while the bulk-loading into and indexing in \rowstore are slow taking $2.5$ and $4.3$ hours respectively.
Then loading the indexed data from \rowstore into \columnstore takes an additional $7$ hours.
These large loading (and indexing) times are due to the billions of rows in the \sql{reception_edges_denorm} table as seen in \Cref{tab:reception-edges-denorm-stats}.
The above trend is reversed for the top quotes task, where materialization in \spark is slow, taking $24$ minutes, while the bulk-loading into and indexing in \rowstore is fast, taking $3.5$ and $7.1$ minutes respectively.
Furthermore, loading from \rowstore into \columnstore takes only an additional $5$ minutes.
This behavior is due to the computational effort required to calculate the \nworks statistic and the relatively small size of the resulting \sql{source_piece_statistics_denorm} table as seen in \Cref{tab:source-piece-statistics-denorm-stats}.


 
In summary, the materialization practicalities discussed above indicate that the \spark framework is essential to materialize the additional data required for the system.
Furthermore, the discussion of loading practicalities highlight the limitations of the \columnstore framework to bulk-load data both after pre-processing and materialization.
Lastly, the results show the overhead of indexing the tables in the \rowstore framework after bulk-loading from \spark. 

\subsection{Latency results}
\label{sec:workload-results}
We present the query latency results for each downstream task across different workloads.

\begin{figure*}
    \centering
    \begin{subfigure}{\linewidth}
        \centering
        \includegraphics[width=0.9\textwidth]{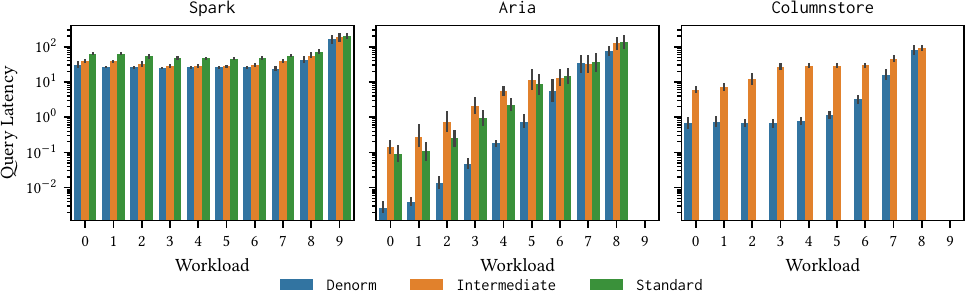}
        \caption{\hpchd dataset}
        \label{fig:hpc-hd-reception-duration}
    \end{subfigure}
    \begin{subfigure}{\linewidth}
        \centering
        \includegraphics[width=0.9\textwidth]{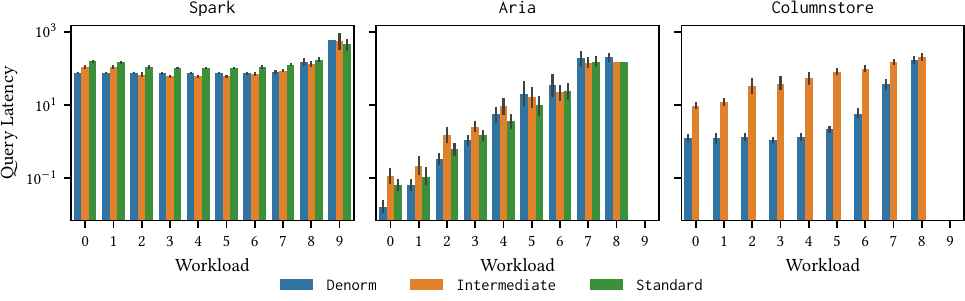}
        \caption{\hpchdnewspapers dataset}
        \label{fig:hpc-hd-newspapers-reception-duration}
    \end{subfigure}
    \caption{Reception query latency for different workloads, frameworks and datasets. Missing results indicate that the query didn't finish after 5 minutes.}
    \label{fig:reception-workload-duration}
\end{figure*}

\spara{Reception Task Results.}
\Cref{fig:reception-workload-duration} presents the latency results for the reception task.
In each plot, the x-axis corresponds to the different query workloads and the colors correspond to the different normalization levels.

From the results, we see that the \standard query in the \columnstore framework never finishes, even for the smallest workload.
Furthermore, none of the queries finish for the highest workload in both \columnstore and \rowstore.
In contrast, \spark has a high but consistent query latency across all workloads.

At lower workloads, the \denorm query has the lowest latency across all frameworks.
Specifically, the \rowstore framework has the lowest latency compared to the other frameworks due to the presence of indexed tablesx.
However, at higher workloads ($6,7$ and $8$), the latency of \rowstore is dominated by the iteration over the large result set and not by the query execution.
This is evidenced by the similar query latency for all three normalization levels.
Furthermore, the latency for all frameworks at higher workloads are similar.

\begin{figure}
    \centering
    \includegraphics[width=0.75\linewidth]{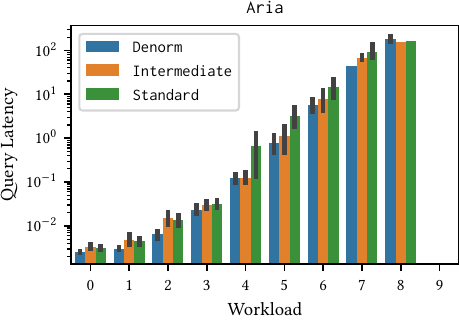}
    \caption{Reception query latency for \rowstore framework with hot cache for \hpchdnewspapers dataset}
    \label{fig:hpc-hd-newspapers-reception--hot-cache-duration}
\end{figure}

Interestingly, the \intermediate query in the \rowstore framework has a higher latency compared to the \standard query.
Furthermore, the \denorm query also has a larger latency at higher workloads in the \hpchdnewspapers dataset.
These higher latencies were due to system page faults stemming from a cold cache.
Therefore, to heat up the cache, we re-ran each query twice and recorded the latency from the second run.
We present these hot-cache results for the \hpchdnewspapers dataset in \Cref{fig:hpc-hd-newspapers-reception--hot-cache-duration}.
From these results, we observe that all queries have an overall lower query latency.
Moreover, we now see that the \intermediate and \denorm queries perform similarly with latency lower than the \standard query.

\begin{figure*}
    \centering
    \begin{subfigure}{\linewidth}
        \centering
        \includegraphics[width=0.9\textwidth]{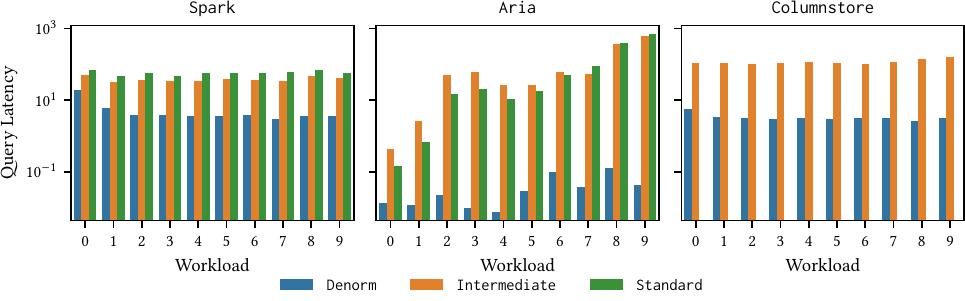}
        \caption{\hpchd dataset}
        \label{fig:hpc-hd-quote-duration}
    \end{subfigure}
    \begin{subfigure}{\linewidth}
        \centering
        \includegraphics[width=0.9\textwidth]{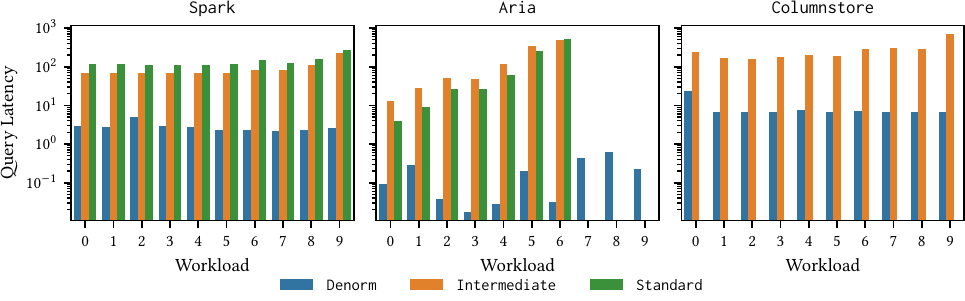}
        \caption{\hpchdnewspapers dataset}
        \label{fig:hpc-hd-newspapers-quote-duration}
    \end{subfigure}
    \caption{Top Quotes query latency for different workloads, frameworks and datasets. Missing results indicate that the query didn't finish after 15 minutes.}
    \label{fig:quote-workload-duration}
\end{figure*}

\spara{Top Quotes Results.}
\Cref{fig:quote-workload-duration} presents the query latency results for the top quotes task.
Similar to the reception task, we observe that the \denorm query has the lowest latency across all frameworks. 
However, the gap in query latency to the \standard and \intermediate queries are much larger here due to the cost to compute the \nworks statistic.
Additionally, the latency gap in the \rowstore framework is two orders of magnitude due to the presence of indexed tables with pre-computed \nworks statistics.

However, the \intermediate and \standard queries on the \rowstore framework fail at higher workloads for the \hpchdnewspapers dataset.
This is due to the larger aggregations required to compute the \nworks statistic at higher workloads.
In contrast, the \intermediate query in the \columnstore framework has a consistent query latency.
This consistency is due to the nature of the columnar data storage and execution engine built for aggregation queries.
For similar reasons, the \spark framework also has a consistent latency for both the \intermediate and \standard queries across all workloads.

\begin{figure*}
    \centering
    \includegraphics[width=0.8\linewidth]{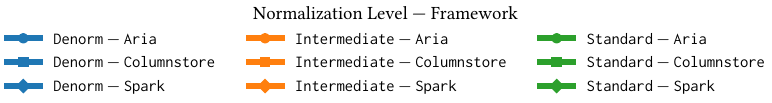}

    \begin{subfigure}{0.45\linewidth}
        \centering
        \includegraphics[width=\textwidth]{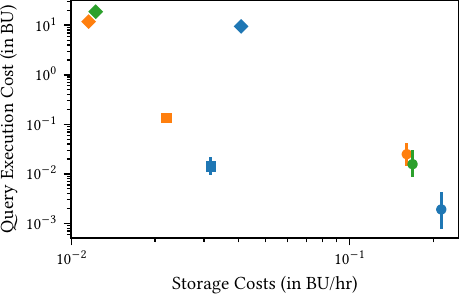}
        \caption{\hpchd dataset}
        \label{fig:hpc-hd-reception-costs-trade-off}
    \end{subfigure}%
    \begin{subfigure}{0.45\linewidth}
        \centering
        \includegraphics[width=\textwidth]{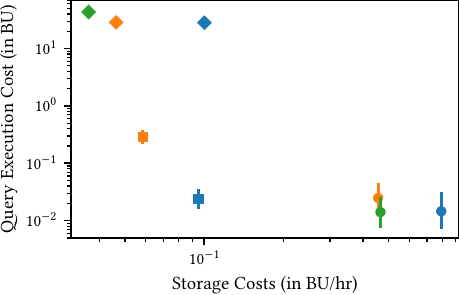}
        \caption{\hpchdnewspapers dataset}
        \label{fig:hpc-hd-newspapers-reception-cost-trade-off}
    \end{subfigure}
    \caption{Trade-off plots between query execution and data storage costs for the Reception task. Each framework has a different marker and each normalization levels has a different color. Lower is better for both metrics.}
    \label{fig:reception-cost-trade-offs}
\end{figure*}

\begin{figure*}
    \centering
    \begin{subfigure}{0.45\linewidth}
        \centering
        \includegraphics[width=\textwidth]{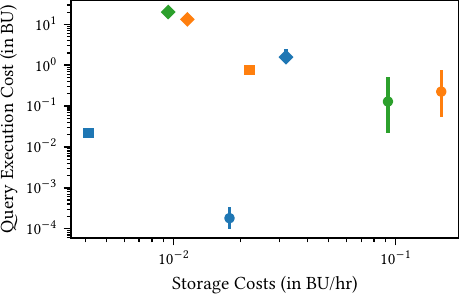}
        \caption{\hpchd dataset}
        \label{fig:hpc-hd-quote-cost-trade-off}
    \end{subfigure}%
    \begin{subfigure}{0.45\linewidth}
        \centering
        \includegraphics[width=\textwidth]{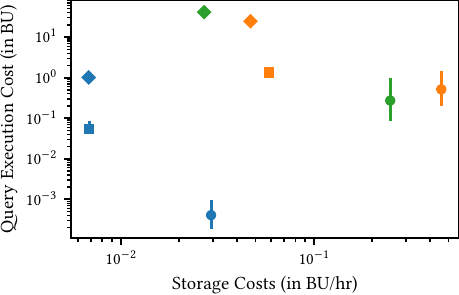}
        \caption{\hpchdnewspapers dataset}
        \label{fig:hpc-hd-newspapers-quote-cost-trade-off}
    \end{subfigure}
    \caption{Trade-off plots between query execution and data storage costs for the Top Quotes task. Each framework has a different marker and each normalization levels has a different color. Lower is better for both metrics.}
    \label{fig:quote-cost-trade-offs}
\end{figure*}

\subsection{Data Storage and Execution Cost Results}
\label{sec:cost-results}
Using the size on disk of the materialized data for each normalization level and the duration of each downstream query, we compute the data storage and query execution costs for a framework using the rates shown in \Cref{tab:storage-execution-costs}
For both downstream tasks, we omit the highest query workload as they do not reflect real-world useworkloads of the system.
We report these costs as trade-off plots in \Cref{fig:reception-cost-trade-offs} and \Cref{fig:quote-cost-trade-offs} for the reception and top quotes tasks respectively.
In each plot, the y-axis is the query execution cost measured in CSC Business Units (BU) and the x-axis is the data storage cost measured in billing units per hour (BU/hr).
The variance in the query execution cost is caused due to the range of workloads evaluated.  
The desirable region for a design choice is the bottom-left which corresponds to both low storage and query execution costs.

Let us first analyze the reception task results shown in \Cref{fig:reception-cost-trade-offs}.
We observe that \rowstore has the lowest execution cost due to its low latency.
However, it also has the highest storage cost due to the large materialized \sql{reception_edges_denorm} table and its index.
In contrast, \spark has the lowest storage costs and the highest execution costs.
Finally, the \columnstore framework with \denorm level data offers a good trade-off between the two costs, especially for the \hpchdnewspapers dataset.

Next, we analyze the top quotes task results shown in \Cref{fig:quote-cost-trade-offs}.
In this task, the smaller size of the \denorm data and the lower latency from indexed tables results in the \rowstore framework habing the lowest query execution cost.
However, at the \intermediate and \standard levels, the data storage costs of the \rowstore framework is the largest.
Moreover, the \rowstore framework fails at higher workloads for \hpchdnewspapers dataset as shown in \Cref{fig:hpc-hd-newspapers-quote-duration}.
Therefore, the \columnstore framework is a better choice for the \intermediate normalization level and the \spark framework offers the best trade-off for the \standard normalization level.

\subsection{Discussion}
\label{sec:discussion}

We have three main takeaways from the results presented in the above section.

\spara{\denorm queries have the lowest latency}
For both the downstream tasks and across frameworks, the \denorm query has the lowest query latency.
The materialization of the task-specific data greatly helps to lower the time to get answers to queries, especially at lower workloads.
However, for the reception task, the \denorm level comes with a significant materialization and data storage cost.
In contrast, for the top quotes task, both the materialization and data storage costs are lower, making \denorm normalization level the best option.

\spara{\spark is irreplaceable and good at higher workloads} 
As shown in \Cref{sec:post-processing-and-loading-results}, \spark is the only viable approach for materialization and loading the data required for the \intermediate and \denorm normalization levels.
The \spark framework is also the only framework that works at the highest workload for both downstream tasks.
Additionally, even at higher workloads, the framework offers a consistent, yet high, query latency even with the \standard data normalization level.
%
%
%

\spara{\columnstore offers a good trade-off}
The \columnstore framework is the compromise between the \spark and \rowstore frameworks.
It has higher latency when compared to \rowstore, but is consistent across workloads.
And similar to \spark, the compressed columnar storage of the \columnstore framework results in very low storage costs, even with large \denorm level materialized data. 
The two main drawbacks are inability to run any \standard level query and the significantly high materialization and loading costs.
%
\section{Related Works}
\label{sec:related-works}

There are several systems for identifying and analyzing text reuses from in recent literature \cite{during2023ImpressoTextReuse,harris2018FindingParallelPassages,vesanto-etal-2017-system,roe2016digging}.
These systems use different approaches to identifying text reuses from noisy OCR such as n-grams~\cite{smith2015ComputationalMethodsUncovering}.
All these systems also have a pre-processing phase that includes filtering, merging and clustering the identified text reuses.
However, many of these systems focus primarily on historical newspaper corpora, and with at most a couple million text reuses.
In contrast, our \system system is built to work with billions of text reuses from a mixture of historical books and newspaper corpora.
Furthermore, we evaluate our system on complex downstream analytical tasks while the other systems simply provide a front-end interface of the pre-processed text reuses.

Regarding the \rowstore and \columnstore frameworks considered in this paper, there are several studies comparing row store and column store database management systems \cite{abadi2008ColumnstoresVsRowstores,chaalal2021FindingBestColumn}.
However, these studies are generic using TCP-H benchmarks to analyze the trade-off between different storage and execution engines.
In contrast, we provide a detailed evaluation of the database frameworks along with the \spark framework.
Moreover, we discuss the materialization costs and the trade-offs between the different frameworks in context of analyzing large-scale text reuse data.
\section{Conclusion}
\label{sec:conclusion}
In this paper, we presented \system, our system for analyzing text reuses in large historical corpora.
We studied the impact of data management, data storage and query execution on the performance of the system for the two downstream analysis tasks of reception and top quotes.
For data management, we considered three levels of data normalization, namely, \standard, \intermediate and \denorm, each has its own data materialization and task-related queries.
For data storage and query execution, we considered three frameworks:  \spark, \rowstore and  \columnstore.
We analyzed the trade-off of each combination of normalization level and framework in terms of query latency, data storage size and query execution costs.
Our results showed that the \denorm normalization level in the \rowstore framework has the lowest query latency with higher data materialization and storage costs.
We found that the \spark framework is irreplaceable for the large-scale data processing of the system regardless of the high execution cost.
We noted the limitations of the \columnstore framework in terms of data loading costs and incompatibility with the \standard normalization level.
However, at the \denorm level, the \columnstore framework offers a good trade-off between data storage and execution costs, especially as the dataset scales.   
Lastly, our contributions showcased the system's scalability from the \hpchd dataset to the $6\times$ larger \hpchdnewspapers dataset.

In future work, we would like to evaluate the pre-processing pipeline and other execution frameworks.
In particular, a quantitative and qualitative analysis of the text reuse clusters obtained might offer insights into the nature of text reuses detected during pre-processing.
Furthermore, evaluating a graph database framework for the text reuse network would be interesting.

In conclusion, our paper represents a significant advance in evaluating and optimizing scalable data science system for humanities research.  
Our findings provide valuable insights into the trade-offs offered by each design choices towards the practical deployment of the system.



\bibliographystyle{ACM-Reference-Format}
\bibliography{references}

\end{document}